\documentclass[3p,twocolumn,sort&compress,fleqn]{elsarticle}

\usepackage{amsmath}
\usepackage{amssymb}
\usepackage{bm}
\usepackage{braket}
\usepackage{cprotect}
\usepackage{lmodern}
\usepackage{ulem}
\usepackage{listings,jvlisting}
\usepackage{physics}
\usepackage{xspace}
\usepackage{booktabs}
\usepackage{tikz}
\usetikzlibrary{shapes.geometric,arrows}

\journal{Computer Physics Communications}

\newcommand{\tre}[1]{#1}

\def\vec#1{\boldsymbol #1}

\usepackage{natbib}
\usepackage{hyperref}
\hypersetup{
        colorlinks=true,
}

\long\def\beginmypgfpdfnamed#1#2\endmypgfpdfnamed{\includegraphics{#1}}

\graphicspath{{./pyfig/}}

\newcommand{\hwave}{\textsf{H-wave}\xspace}
\newcommand{\HPhi}{$\mathcal{H}\Phi$\xspace}
\newcommand{\mVMC}{mVMC\xspace}

\begin{document}
\begin{frontmatter}

\title{\hwave{} -- A Python package for the Hartree-Fock approximation and the random phase approximation }

\author[issp]{Tatsumi Aoyama}\corref{author}\ead{aoym@issp.u-tokyo.ac.jp}
\author[issp]{Kazuyoshi Yoshimi}\ead{k-yoshimi@issp.u-tokyo.ac.jp}
\author[issp]{Kota Ido}
\author[issp]{Yuichi Motoyama}
\author[nagoya]{Taiki Kawamura}
\author[issp]{Takahiro Misawa}
\author[issp]{Takeo Kato}
\author[nagoya]{Akito Kobayashi}

\cortext[author]{Corresponding author.}
\address[issp]{Institute for Solid State Physics, University of Tokyo, Chiba 277-8581, Japan}
\address[nagoya]{Department of Physics, Nagoya University, Nagoya 464-8602, Japan}

\begin{abstract}
\hwave is an open-source software package for performing the Hartree--Fock approximation (HFA) and random phase approximation (RPA) for a wide range of Hamiltonians of interacting fermionic systems. 
In HFA calculations, \hwave examines the stability of several symmetry-broken phases, such as anti-ferromagnetic and charge-ordered phases, in the given Hamiltonians at zero and finite temperatures.
Furthermore, \hwave calculates the dynamical susceptibilities using RPA to examine the instability toward the symmetry-broken phases. 
By preparing a simple input file for specifying the Hamiltonians, users can perform HFA and RPA for standard Hamiltonians in condensed matter physics, such as the Hubbard model and its extensions.
Additionally, users can use a \textsc{Wannier90}-like format to specify fermionic Hamiltonians.
A \textsc{Wannier90} format is implemented in \textsc{RESPACK} to derive 
{\it ab initio} Hamiltonians for solids. HFA and RPA for the {\it ab initio} Hamiltonians
can be easily performed using \hwave.
In this paper, we first explain the basis of HFA and RPA, and 
the basic usage of \hwave, including download and installation.
Thereafter, the input file formats implemented in \hwave, including the \textsc{Wannier90}-like format for specifying the interacting fermionic Hamiltonians, are discussed.
Finally, we present several examples of \hwave such as zero-temperature HFA calculations for the extended Hubbard model on a square lattice, finite-temperature HFA calculations for the Hubbard model on a cubic lattice, and RPA in the extended Hubbard model on a square lattice.
\end{abstract}


\end{frontmatter}

{\bf PROGRAM SUMMARY}

\begin{small}
\noindent
{\em Program Title:} \hwave \\
{\em CPC Library link to program files:} \\
{\em Developer's repository link:} \url{https://github.com/issp-center-dev/H-wave} \\
{\em Code Ocean capsule:} \\
{\em Licensing provisions:} GNU General Public License version 3 \\
{\em Programming language:} Python3 \\
{\em External routines/libraries:} NumPy, SciPy, Tomli, Requests. \\
{\em Nature of problem:}
Physical properties of strongly correlated electrons are examined such as ground state phase structure and response functions at zero and finite temperatures. \\
{\em Solution method:}
Calculations based on the unrestricted Hartree-Fock approximation and the random phase approximation are performed for the quantum lattice models such as the Hubbard model and its extensions. \\
\end{small}

\section{Introduction} 

Clarifying how electron correlations in solids induce several symmetry-broken phases has long been a crucial issue in the field of condensed matter physics~\cite{Imada1998,Fazekas1999}.
Although deriving tight-binding models from {\it ab initio} calculations is now a standard daily procedure~\cite{Pizzi2020, PhysRevB.70.195104, doi:10.1143/JPSJ.79.112001}, solving the obtained effective models by considering the effects of Coulomb interactions remains as a non-trivial problem. To solve effective models numerically, various numerical algorithms such as the exact diagonalization~\cite{Dagotto1994}, 
quantum Monte Carlo~\cite{Gubernatis2016}, and tensor network~\cite{Schollwock2005,Orus2014,Orus2019} methods have been developed to date ~\cite{Fehske2008,Avella2013}.
However, these algorithms frequently incur high computational cost because they consume a significant amount of memory or execution time.

Numerical algorithms based on the Hartree--Fock approximation (HFA) have long been used in condensed matter physics~\cite{Mahan,Ring}.
Unrestricted HFA (UHFA), which is a simple and fundamental method for treating electronic states in molecules and solids, is used to examine the stability of ordered phases.
For example, the HFA has been used for analyzing the
ordered phases in the Hubbard model~\cite{Hirsch_PRB1985, Penn_PR1966}, spin and orbital orderings in 3$d$ transition-metal oxides~\cite{Mizokawa1996}, and the spin and charge ordered phases in organic compounds~\cite{Seo2004}.
Similarly, random phase approximation (RPA), which is an algorithm for calculating response functions consistent with the UHFA~\cite{Mahan, Fetter, Kadanoff1994}, is used for detecting instability toward ordered phases through the divergence of susceptibility.
In comparison with other sophisticated algorithms, the main advantage of the UHFA and the RPA is low numerical cost.
Therefore, they are suitable for clarifying general trends in ordered phases in correlated electron systems.

Program codes for UHFA and RPA have been implemented and maintained by researchers for specific lattice models.
To our knowledge, there are no widely used packages for UHFA and RPA, except for a supplementary RPA package included in \textsf{triqs}~\cite{triqs,triqsref} and the supplementary real-space UHF package included in \mVMC~\cite{mvmc-manual}.
Recently, HF and RPA analyses combined with first-principles calculations have become popular in treating real materials. 
When handling such complex models derived from the first-principles calculations, it will take much time to modify existing packages or to newly implement programs. 
For an easy application of the UHFA and RPA to such complex models, the development of a software package that treats various effective models of solids in a simple format is desirable.
Recently, \hwave was released as an open-source Python package with a simple and flexible user interface~\cite{Hwave}. 
Using \hwave, users can perform calculations by applying UHFA and RPA to widely studied quantum lattice models by preparing only one input file with less than ten lines. 
\hwave can treat the \textsc{Wannier90}-like format and has interfaces for smooth connection with \mVMC~\cite{mvmc-manual} and \HPhi~\cite{hphi-manual}, which are program packages for more accurate calculation taking into account electron correlations.

In this paper, we introduce \hwave version 1.0.0.
The remainder of this paper is organized as follows.
Models that can be treated using \hwave are explained in Sec.~\ref{Sec:Model}. The basis of the UHFA and RPA is discussed in Secs.~\ref{Sec:UHF} and \ref{Sec:RPA}, respectively.
In Sec.~\ref{Sec:Usage}, the basic usage of \hwave, for example, downloading, installing, and running operations, is explained. The formats of the input files, including the \textsc{Wannier90}-like format for specifying the Hamiltonians, are elaborated.
Three applications of \hwave are demonstrated in Sec.~\ref{Sec:Examples}, such as the ground-state UHFA calculations of the extended Hubbard model on a square lattice, finite-temperature UHFA calculations of the Hubbard model on a cubic lattice, and 
RPA calculations of the extended Hubbard model on a square lattice.
Section~\ref{Sec:Summary} presents a summary of the study.

\section{Model and Algorithm}
\label{Sec:ModelaAlgorithm}

\subsection{Model}
\label{Sec:Model}
In \hwave, the following extended Hubbard model can be treated:
\begin{align}
  {\cal H} &= {\cal H}_0 + {\cal H}_{\rm int}, \label{Hubbard1}\\
  {\cal H}_0 &= \sum_{\langle i\alpha;j\beta \rangle}
        \left(
        t_{ij}^{\alpha \beta}c_{i\alpha}^{\dag} c_{j\beta}^{\mathstrut}
        + \text{H.c.}
        \right), \\
{\cal H}_{\rm Int} &= \sum_{i,j,k,l}\sum_{\alpha,\alpha^\prime,\beta,\beta^\prime}
{\mathcal I}_{ijkl}^{\alpha\alpha^\prime\beta\beta'}
       c_{i\alpha}^{\dagger} c_{j\alpha^\prime} c_{k\beta}^{\dagger} c_{l\beta^\prime},\label{eq:general-ham}
\end{align}
where $i$, $j$, $k$, and $l$ ($\in 0,\cdots, N_\text{site}-1$) denote the lattice points with $N_\text{site}$ being the number of lattice sites, and $\alpha$, $\alpha^\prime$, $\beta$, and $\beta^\prime$ specify the general orbitals including the orbitals and spins in a unit cell. 
$c_{i\alpha}^{\dag}$ ($c_{i\alpha}$) is the creation (annihilation) operator for an electron with a general orbital $\alpha$ at the $i$th site. 
$\langle i\alpha;j\beta \rangle$ and $t_{ij}^{\alpha\beta}$ represent the
bond pair and the hopping integral between sites $(i,\alpha)$ and $(j, \beta)$, respectively. 
${\cal H}_{\rm Int}$ is a general two-body Hamiltonian and the two-body interaction satisfies ${\mathcal I}_{ijkl}^{\alpha \alpha'\beta \beta'} = \left[{\mathcal I}_{lkji}^{\beta'\beta\alpha'\alpha} \right]^{*}$ for satisfying the Hermiticity.

\subsection{Unrestricted Hartree--Fock approximation}
\label{Sec:UHF}
In the unrestricted Hartree-Fock approximation (UHFA),
we can decouple the general two-body interactions as 
\begin{multline}
  c^\dagger_{I} c_{J}^{\mathstrut} c^\dagger_{K} c_{L}^{\mathstrut}
  =
  -c_{I}^{\dagger} c_{K}^{\dagger} c_{J}^{\mathstrut} c_{L}^{\mathstrut}
  + c^\dagger_{I} c_{L}^{\mathstrut} \delta_{J,K} \\
  \sim
  -\langle c_{I}^{\dagger} c_{L}^{\mathstrut} \rangle c_{K}^{\dagger} c_{J}^{\mathstrut}
  - c_{I}^{\dagger} c_{L}^{\mathstrut} \langle c_{K}^{\dagger} c_{J}^{\mathstrut} \rangle
  + \langle c_{I}^{\dagger} c_{J}^{\mathstrut} \rangle c_{K}^{\dagger} c_{L}^{\mathstrut} \\
  + c_{I}^{\dagger} c_{J}^{\mathstrut} \langle c_{K}^{\dagger} c_{L}^{\mathstrut} \rangle
  + \bigl(
  \langle c_{I}^{\dagger} c_{L}^{\mathstrut} \rangle \langle c_{K}^{\dagger} c_{J}^{\mathstrut} \rangle
  - \langle c_{I}^{\dagger} c_{J}^{\mathstrut} \rangle \langle c_{K}^{\dagger} c_{L}^{\mathstrut} \rangle
  \bigr) \\
  + c^\dagger_{I} c_{L}^{\mathstrut} \delta_{J,K},
  \label{eq:decouple}
\end{multline}
where we adopt $I\equiv(i,\alpha)$, $J\equiv(j,\alpha^\prime)$, $K\equiv(k,\beta^\prime)$, and $L\equiv(l,\beta)$ as notations for brevity, and the Kronecker's delta is denoted as $\delta_{J,K}$.
Using the UHFA, the Hamiltonian is generally denoted as 
\begin{equation}
\mathcal{H}_\text{UHFA} = \sum_{I,J} c^\dagger_{I} H_{IJ}^{\mathstrut}c_{J}^{\mathstrut} = \vec{c}^\dagger \vec{H} \vec{c},
\label{eq:uhf:ham}
\end{equation}
where $\vec{H}$ denotes a matrix with elements represented as $H_{IJ}$, 
and $\vec{c}$ ($\vec{c}^{\dagger}$) represents a column (row) vector with elements denoted as $c_{I}$ ($c_{I}^{\dagger}$).
Since $\vec{H}$ is an Hermite matrix, the Hamiltonian can be diagonalized by the unitary matrix $\vec{U}$ as
\begin{align}
\vec{H}_{\text{UHFA}}&=\vec{U} \vec{\Lambda} \vec{U}^\dagger, &
\vec{\Lambda}&=\text{diag}(\lambda_{0},\lambda_{1},\dots)
\label{eq:uhf:diag}
\end{align}
where $\lambda_{n}$ represents the $n$th eigenvalue of $\vec{H}$.
By defining $\vec{d} = \vec{U}^\dagger \vec{c}$, $\mathcal{H}_\text{UHFA}$ can be rewritten as
\begin{align}
  \mathcal{H}_\text{UHFA} &= \vec{d}^\dagger \vec{\Lambda} \vec{d} =  \sum_{n} \lambda_{n} d_n^\dagger d_n.
\end{align}
Thus, the free energy from the UHFA can be expressed as 
\begin{multline}
  F = \mu N_e
  - \frac{1}{\beta}\sum_n \ln \left[ 1+\exp (-\beta(\lambda_{n} - \mu)) \right] \\
  + \sum_{I,J,K,L} {\mathcal I}_{IJKL} \Bigl(
  \langle c_{I}^{\dagger} c_L\rangle \langle c_{K}^{\dagger} c_J\rangle
  - \langle c_{I}^{\dagger} c_J\rangle \langle c_{K}^{\dagger} c_L\rangle
  \Bigr),
\end{multline}
where $\beta$ denotes the inverse temperature, $N_e$ is the total number of particles, and $\mu$ represents the chemical potential. 

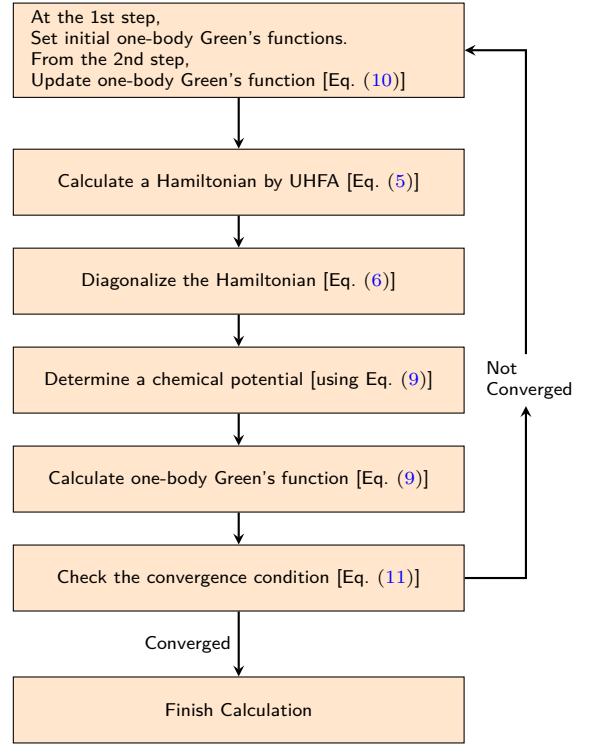
\begin{figure}[bht]
    \centering
  \begingroup
  \scriptsize\sf
  \tikzstyle{process} = [rectangle, minimum width=.35\textwidth, minimum height=8ex, text centered, draw=black, fill=orange!20, text width=.35\textwidth]
  \tikzstyle{arrow} = [thick, ->, >=stealth]
  \begin{tikzpicture}[node distance=12ex]
    \node (init) [process] {\parbox{.95\textwidth}{At the 1st step, \\ Set initial one-body Green's functions. \\ From the 2nd step, \\ Update one-body Green's function [Eq.~\eqref{eq:uhf:update}]}};
    \node (ham)  [process, below of=init, yshift=-4ex] {Calculate a Hamiltonian by UHFA [Eq.~\eqref{eq:uhf:ham}]};
    \node (diag) [process, below of=ham]  {Diagonalize the Hamiltonian [Eq.~\eqref{eq:uhf:diag}]};
    \node (chem) [process, below of=diag] {Determine a chemical potential [using Eq.~\eqref{eq:uhf:calc-green}]};
    \node (calc) [process, below of=chem] {Calculate one-body Green's function [Eq.~\eqref{eq:uhf:calc-green}]};
    \node (chck) [process, below of=calc] {Check the convergence condition [Eq.~\eqref{eq:uhf-residue}]};
    \node (fin)  [process, below of=chck, yshift=-4ex] {Finish Calculation};

    \draw [arrow] (init) -- (ham);
    \draw [arrow] (ham)  -- (diag);
    \draw [arrow] (diag) -- (chem);
    \draw [arrow] (chem) -- (calc);
    \draw [arrow] (calc) -- (chck);
    \draw [arrow] (chck) -- node [anchor=east] {Converged} (fin);
    \node (mid)  [right of=chem, xshift=.15\textwidth, text width=4em] {\parbox{0pt}{Not \\ Converged} };
    \draw [arrow] (chck) -| (mid);
    \draw [arrow] (mid)  |- (init);
  \end{tikzpicture}
  \endgroup
    \caption{Schematic flow of the UHFA calculation.}
    \label{fig:uhf-flow}
\end{figure}

In the UHFA calculations,
we start from the initial one-body Green's functions, 
which are then updated as 
\begin{equation}
\ev*{c_{I}^\dagger c_{J}}^{\mathstrut} = \sum_{n} U_{In}^* U_{Jn}^{\mathstrut} \ev*{d_n^\dagger d_n} = \sum_{n} \frac{U_{In}^* U_{Jn}^{\mathstrut}}{1+e^{\beta(\lambda_{n} -\mu)}}.
\label{eq:uhf:calc-green}
\end{equation}
In the canonical calculation in which the number of particles is fixed to \tre{$N_e$}, $\mu$ is determined to satisfy the relation
\tre{$N_e = \sum_{I} \langle c_I^{\dagger} c_I^{\mathstrut} \rangle$} at every step.
In \hwave, the simple-mixing algorithm is employed to update the 
one-body Green's functions, which are defined as
\begin{equation}
\ev*{c_{I}^\dagger c_{J}^{\mathstrut}}^{(n+1)} \leftarrow (1-\alpha_{\rm mix}) \ev*{c_{I}^\dagger c_{J}^{\mathstrut}}^{(n)} +  \alpha_{\rm mix} \ev*{c_{I}^\dagger c_{J}^{\mathstrut}}^{(n+1)},
\label{eq:uhf:update}
\end{equation}
where $\ev*{c_{I}^\dagger c_{J}^{\mathstrut}}^{(n)}$ represents the
Green's functions at the $n$th step and
$\alpha_{\rm mix}$ denotes a parameter between $0$ and $1$.
The iterations are repeated until the convergence condition is satisfied in which the residue $R$ becomes smaller than a given criterion $\varepsilon$, such that
\begin{equation}
  R = \dfrac{1}{2N_\text{site}^2}
  \sqrt{\sum_{IJ} \left|
    \langle c_I^\dagger c_J^{\mathstrut} \rangle^{(n+1)}
    -
    \langle c_I^\dagger c_J^{\mathstrut} \rangle^{(n)}
    \right|^2} < \varepsilon .
  \label{eq:uhf-residue}
\end{equation}
The calculation is unsuccessful or fails if the convergence condition is not met within a specified number of iterations. 
In Fig.~\ref{fig:uhf-flow}, a schematic flow 
of the calculation of the UHFA calculation is shown for convenience.

\subsubsection{Hartree--Fock approximation in the momentum space}
When a two-body Hamiltonian satisfies the translational symmetry, 
we can efficiently perform the HFA calculation in the momentum space.
In the following, the translational-invariant two-body Hamiltonian below is considered instead of Eq.~\eqref{eq:general-ham}
\begin{multline}
  {\cal H}_{\rm Int} \equiv \sum_{ij}\sum_{\alpha,\alpha^\prime,\beta,\beta^\prime}
W_{ij}^{\alpha\alpha^\prime\beta\beta^\prime}
        c_{i\alpha}^{\dagger} c_{i\alpha^\prime}^{\mathstrut} c_{j\beta}^{\dagger} c_{j\beta^\prime}^{\mathstrut},
        \label{eq:hk-genearal}
\end{multline}
where a two-body interactions satisfies $W_{ij}^{\alpha\alpha^\prime\beta\beta^\prime} = \left[W_{ji}^{\beta^\prime\beta\alpha'\alpha}\right]^*$ for the Hermiticity.
The Fourier transformations of the operator are defined as
\begin{align}
c_{\vec{r}_{i}\alpha}           &= \frac{1}{\sqrt{N_\text{site}}} \sum_{\vec{k}} e^{i\vec{k} \cdot \vec{r}_{i}}c_{\vec{k}\alpha},\\
c_{\vec{r}_{i}\alpha}^{\dagger} &= \frac{1}{\sqrt{N_\text{site}}} \sum_{\vec{k}} e^{-i\vec{k} \cdot \vec{r}_{i}}c_{\vec{k}\alpha}^{\dagger},  
\end{align}
where $N_\text{site}$ denotes the number of the sites and $\vec{r}_{i}$ is a position vector at the $i$th site. 
The one-body Hamiltonian in the real space is rewritten as 
\begin{align}
h_0 &=
\sum_{\vec{k},\alpha,\beta}
c_{\vec{k}\alpha}^\dagger h_{\alpha\beta}(\vec{k}) c_{\vec{k}\beta}^{\phantom\dagger}.
\end{align}
Since the system has a translational symmetry, the coefficients depend only on the translation vectors, $\vec{r}_{ij}=\vec{r}_j - \vec{r}_i$ and
$h_{\alpha\beta}(\vec{k})=\sum_{j}{\tilde t}_{0j}^{\alpha\beta}e^{-i\vec{k}\cdot\vec{r}_{0j}}$, where ${\tilde t}_{0j}^{\alpha\beta}$ includes $t_{0j}^{\alpha\beta}$ and the decoupled two-body interaction terms.
As the Hamiltonian is diagonal with respect to the wave number $\vec{k}$, 
the eigenvalue and eigenvector calculations are simplified from the diagonalization of the matrix of the size
$N_\text{site}N_\text{orbit} \times N_\text{site}N_\text{orbit}$
to that of $N_\text{site}$ matrices of the size
$N_\text{orbit} \times N_\text{orbit}$, where $N_\text{orbit}$ denotes the number of orbitals, including the spin degrees of freedom. Thus, the calculation costs are reduced.

\subsection{Random Phase Approximation}
\label{Sec:RPA}
The random phase approximation (RPA) is used for calculating the response functions, such as charge/spin susceptibility.
By examining the temperature dependence of the charge/spin susceptibility, the instability of charge/spin ordered phases can be detected.
In \hwave, RPA can be implemented in momentum space and users can obtain the dynamical susceptibility $\chi(\vec{q},i\nu_{n})$, where $\nu_{n}$ represents the bosonic Matsubara frequency, $\nu_{n}=2n\pi k_B T$ $(n = -N_{\omega}/2, \cdots, N_{\omega}/2-1)$ with the even cutoff integer $N_{\omega}$. Here, $T$ is the temperature and $k_B$ is the Boltzmann constant, and for simplicity, we set $k_B = 1$ in the following.
We note that it is necessary to perform the analytical continuation to obtaining the real-frequency dynamical susceptibility $\chi(\vec{q},\omega)$, which can be measured by experiments.
In this section, we briefly introduce the method for calculating the dynamical susceptibility using the RPA.

By applying Fourier transformation to the two-body Hamiltonian ${\cal H}_{\rm Int}$ defined in Eq.~\eqref{eq:hk-genearal}, 
the Hamiltonian can be rewritten as
\begin{multline}
  {\cal H}_{\rm Int} =
   \frac{1}{2N_{\rm site}} \sum_{{\bm k},{\bm k}^\prime,{\bm q}} \sum_{\alpha,\alpha^\prime,\beta,\beta^\prime}
W^{\alpha\alpha^\prime\beta\beta^\prime}_{{\bm q}}
  \\ \times
  c_{{\bm k}+{\bm q},\alpha}^{\dagger}
  c_{{\bm k},\alpha^\prime}^{\mathstrut}
  c_{{\bm k}^\prime-{\bm q},\beta}^{\dagger}
  c_{{\bm k}^\prime,\beta^\prime}^{\mathstrut}.
\end{multline}
In the RPA, the scattering processes by the interaction are considered on the basis of eigenvectors of the non-interacting Hamiltonian ${\cal H}_0$. 
On this basis, the two-body interaction operator can be approximated as
\begin{multline}
c_{\bm{k}+\bm{q},\alpha}^{\dagger}\,
c_{\bm{k},\alpha^\prime}^{\mathstrut}\,
c_{\bm{k}^\prime-\bm{q},\beta}^{\dagger}\,
c_{\bm{k}^\prime,\beta^\prime}^{\mathstrut} \\
\sim \sum_{\gamma, \gamma^\prime}
u_{\alpha \gamma, \bm{k}+\bm{q}}^*\,
u_{\alpha^\prime \gamma, \bm{k}}\,
u_{\beta \gamma^\prime, \bm{k}^\prime-\bm{q}}^*\,
u_{\beta^\prime  \gamma^\prime, \bm{k}^\prime} \\ 
\times
  d_{\bm{k}+\bm{q},\gamma}^{\dagger}
  d_{\bm{k},\gamma}^{\mathstrut}
  d_{\bm{k}^\prime-\bm{q},\gamma^\prime}^{\dagger}
  d_{\bm{k}^\prime,\gamma^\prime}^{\mathstrut},     
\end{multline}
where $c_{\bm{k},\alpha} = \sum_{\gamma} u_{\alpha \gamma, \bm{k}} d_{\bm{k}, \gamma}$,
and $d_{\bm{k}, \gamma}$ denotes the annihilation operator that diagonalizes ${\cal H}_0$. ($\gamma$ denotes the eigenvalue index.) 
Then, the irreducible one-body Green's function can be expressed as
\begin{align}
 G^{(0)\alpha\beta}_{\gamma}({\bm k}, i{\omega_{n}})=
  \frac{u^{\alpha\gamma}({\bm k})u^{*\beta\gamma}({\bm k})}{i{\omega_{n}}-\xi^{\gamma}({\bm k})+\mu},
\label{eq:rpa:green0}
\end{align}
where $\omega_n$ represents the fermionic Matsubara frequency {$\omega_n = (2n+1) \pi T$}.
Generally, the irreducible susceptibility in non-interacting systems can be expressed as
\begin{multline}
  \chi^{(0)\alpha\alpha^\prime, \beta^\prime\beta}({\bm q},i\nu_n)
  = \\
  -\frac{T}{N_\text{site}}
  \sum_{\gamma=1}^{N_{\rm orbit}}\sum_{{\bm k},m}
  G^{(0)\alpha\beta}_{\gamma}({\bm k}+{\bm q}, i{\omega_{m}}+i\nu_n) \, \\ \times
  G^{(0)\beta^\prime\alpha^\prime}_{\gamma}({\bm k}, i\omega_m).
\label{eq:rpa:chi0q}
\end{multline}
By using the irreducible susceptibility, 
the dynamical susceptibility from the RPA, $\chi^{\alpha\alpha^\prime, \beta^\prime\beta}(\vec{q}, i\nu_n)$, can be obtained as
\begin{multline}
  \chi^{\alpha\alpha^\prime, \beta^\prime\beta}(q)
  =
  \chi^{(0)\alpha\alpha^\prime, \beta^\prime\beta}(q) \\
  - \sum_{\alpha_1^\prime\beta_1^\prime}
  \chi^{(0)\alpha\alpha^\prime, \beta_1^\prime\beta_1}(q) \,
  W^{ \alpha_1\alpha_1^\prime\beta_1^\prime\beta_1}_{\bm q} \,
  \chi^{\alpha_1 \alpha_1^\prime , \beta^\prime\beta}(q),
\label{eq:xq}
\end{multline}
where we define $ q \equiv ({\bm q}, i\nu_{n})$ for simplicity. 
Here, the index $\alpha$ includes the orbital and spin degrees of freedom.
 By combining indices into one index, for example, $a=\alpha\times\alpha^\prime$, Eq.~\eqref{eq:xq} can be rewritten in a matrix form as
\begin{align}
  \vec{\chi}(q)
  &=
  \vec{\chi}^{(0)}(q)-\vec{\chi}^{(0)}(q)\vec{W}(q)\vec{\chi}(q) \nonumber\\
  &=
  \left[\vec{I}+\vec{\chi}^{(0)}(q)\vec{W}(q)\right]^{-1}\vec{\chi}^{(0)}(q),
\label{eq:rpa:chiq}
\end{align}
where $I$ denotes the identity matrix.
Here, for describing the equation in matrix notation, we define the two-body interaction $[\vec{W}(q)]^{ab} \equiv W^{ba}_{\bm q}$.
The dimensions of the susceptibilities $\chi^{(0)\alpha\alpha^\prime\beta^\prime\beta}({\bm q},i\nu_n)$ and $\chi^{\alpha\alpha^\prime, \beta^\prime\beta}({\bm q},i\nu_n)$  are given by $N_{\rm orbit}^4 N_{\rm site} N_{\omega}$.

The size of the multidimensional array of susceptibilities can be reduced by separating orbitals and spins.
When the spin-orbital coupling does not exist and ${\cal H}$ does not include terms that mix spin and orbital, the orbital and spin Hilbert spaces become independent.
In this case, 
the two-body interaction can be rewritten as
\begin{equation}
W^{\alpha_{\sigma\sigma^\prime}\beta_{\sigma_1\sigma_1^\prime}}_{\vec{q}}c_{\bm{k}+\bm{q},\alpha \sigma}^{\dagger}c_{\bm{k},\alpha \sigma^\prime}^{\mathstrut}
c_{\bm{k}^\prime-\bm{q},\beta\sigma_1}^{\dagger} c_{\bm{k}^\prime,\beta\sigma_1^\prime}^{\mathstrut},
\end{equation}
where $\alpha_{\sigma\sigma^\prime} \equiv \alpha \sigma \alpha \sigma^{\prime}$ and $\beta_{\sigma_1\sigma_1^\prime}\equiv \beta\sigma_1\beta\sigma_1^\prime$.
Since the scattering processes occur only on the same diagonalized general orbitals, the irreducible susceptibility can be expressed as 
\begin{multline}
  \chi^{(0)\alpha, \beta}_{\sigma\sigma^\prime\sigma_1^\prime\sigma_1}({\bm q},i\nu_n)
  = \\
  -\frac{T}{N_\text{site}}
  \sum_{\gamma=1}^{N_{\rm orb}}\sum_{{\bm k},{m}}
  G^{(0)\alpha\beta}_{\sigma\sigma_1^\prime, \gamma}({\bm k}+{\bm q}, i{\omega_m}+ i{\nu_{n}}) \,
  \\ \times
  G^{(0)\beta\alpha}_{\sigma_1\sigma^\prime, \gamma}({\bm k}, i{\omega_{m}}).
  \label{eq:chi0red}
\end{multline}
Here, the array size can be  reduced to $N_{\rm orb}^2 N_{\rm spin}^4 N_{\rm site} N_{\omega}$ where $N_{\rm orb}$ and $N_{\rm spin}$ denote the number of the orbital and spin degrees of freedom, respectively.
Then, the susceptibility can be obtained as 
\begin{multline}
  \chi^{\alpha, \beta}_{\sigma\sigma^\prime\sigma_1^\prime\sigma_1}(q)
  = 
  \chi^{(0)\alpha, \beta}_{\sigma^{\mathstrut}\sigma^\prime\sigma_1^\prime\sigma_1^{\mathstrut}}(q)
  \\
  - \sum_{\alpha_2^{\mathstrut}\alpha_3^{\mathstrut}}\sum_{\sigma_2^{\mathstrut}, \sigma_2^\prime}\sum_{\sigma_3^{\mathstrut}, \sigma_3^\prime}
  \chi^{(0)\alpha, \alpha_2}_{\sigma\sigma^\prime\sigma_2^\prime\sigma_2}(q) \,
  \\ \times
  W^{{\alpha_3}_{\sigma_3^{\mathstrut}\sigma_3^\prime} {\alpha_2}_{\sigma_2^\prime\sigma_2^{\mathstrut}}}_{\vec{q}}\,
  \chi^{\alpha_3, \beta}_{\sigma_3^{\mathstrut}\sigma_3^\prime,\sigma_1^\prime\sigma_1^{\mathstrut}}(q).
\end{multline}
Like as deriving Eq.~\eqref{eq:xq}, if $\alpha_{\sigma\sigma^\prime}$ is represented as a single index, it can be inserted into a matrix form, and for generalized orbitals it can be denoted as
\begin{align}
  \vec{\chi}(q)
  &=
  \vec{\chi}^{(0)}(q)-\vec{\chi}^{(0)}(q)\vec{W}(q)\vec{\chi}(q) 
  \nonumber\\
  &=
  \left[\vec{I}+\vec{\chi}^{(0)}(q)\vec{W}(q)\right]^{-1}\vec{\chi}^{(0)}(q).
  \label{eq:RPA}
\end{align}

\begin{figure}[htb]
    \centering
  \begingroup
  \scriptsize\sf
  \tikzstyle{process} = [rectangle, minimum width=.35\textwidth, minimum height=8ex, text centered, draw=black, fill=green!20, text width=.35\textwidth]
  \tikzstyle{arrow} = [thick, ->, >=stealth]
  \begin{tikzpicture}[node distance=12ex]
    \node (init) [process] {(Option) Set one-body Green's functions};
    \node (diag) [process, below of=init] {Calculate and diagonalize a one-body Hamiltonian};
    \node (chem) [process, below of=diag] {Determine a chemical potential [using Eq.~\eqref{eq:uhf:calc-green}]};
    \node (calc) [process, below of=chem] {Calculate one-body Green's function [Eq.~\eqref{eq:rpa:green0}]};
    \node (chi0) [process, below of=calc] {Calculate irreducible susceptibility [Eq.~\eqref{eq:rpa:chi0q}]};
    \node (chiq) [process, below of=chi0] {Calculate susceptibility matrix [Eq.~\eqref{eq:rpa:chiq}]};
    \draw [arrow] (init) -- (diag);
    \draw [arrow] (diag) -- (chem);
    \draw [arrow] (chem) -- (calc);
    \draw [arrow] (calc) -- (chi0);
    \draw [arrow] (chi0) -- (chiq);
  \end{tikzpicture}
  \endgroup
    \caption{Schematic flow of the RPA calculation.}
    \label{fig:rpa-flow}
\end{figure}
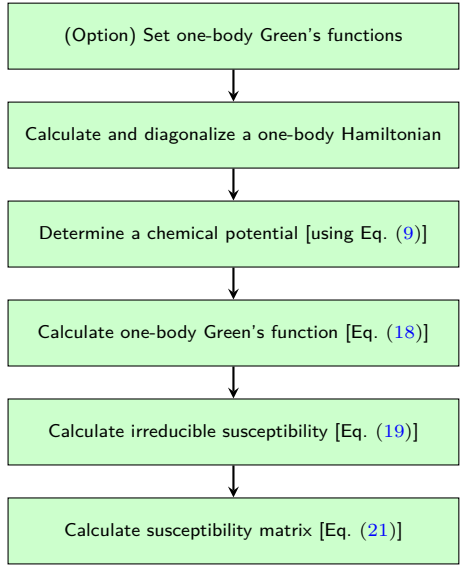

In the above formula, 
it is necessary to store $G^{(0)\beta\alpha}_{\sigma_1\sigma^\prime, \gamma}({\bm k}, i\omega_{n})$ and align the indices of $\gamma$ for $G^{(0)\beta\alpha}_{\sigma_1\sigma^\prime, \gamma}({\bm k}, i\omega_{n})$ before the summation of $\gamma$. However, these procedures increase the numerical cost.
To reduce the numerical cost, in previous studies~\cite{Kobayashi2004a,Takimoto2004,Kubo2006,yoshimi2007,kobayashi2008,Graser2009,kontani2011,Altmeyer2016}
the one-body Green's function is simply calculated as 
$G^{(0)\alpha\beta}_{\sigma\sigma^\prime}({\bm k}, i\omega_{n}) = \sum_{\gamma=1}^{N_{\rm orb}} G^{(0)\alpha\beta}_{\sigma\sigma^\prime, \gamma}({\bm k}, i\omega_{n})$.
Therefore, the irreducible susceptibility is calculated as
\begin{multline}
  \chi^{(0)\alpha, \beta}_{\sigma\sigma^\prime\sigma_1\sigma_1^\prime}({\bm q},i\nu_n)
  = \\
  - \frac{T}{N_\text{site}}
  \sum_{{\bm k},m}
  G^{(0)\alpha\beta}_{\sigma\sigma_1^\prime}({\bm k}+{\bm q}, i\nu_n+ i\omega_m) \, \\ \times
G^{(0)\beta\alpha}_{\sigma_1\sigma^\prime}({\bm k}, i\omega_m).
\end{multline}
Since additional terms, $G^{(0)\alpha\beta}_{\sigma\sigma_1^\prime, \gamma}({\bm k}+{\bm q}, i\nu_m+ i\omega_{n})
G^{(0)\beta\alpha}_{\sigma_1\sigma^\prime, \gamma'} ({\bm k}, i\omega_{n})$ for $\gamma\neq\gamma'$ are included, it can lead to quantitatively different results.
Although this treatment is an \textit{approximation} of the original RPA equations
defined in Eq.~\eqref{eq:RPA}, we adopted this approach. 
We note that when this approximation is applied to orbitals that are clearly independent, a qualitative difference appears because it takes into account the hybridization of orbitals that do not originally exist. After performing the block diagonalization, it is expected to closely agree with the exact RPA when only specific modes are essential, such as near the transition point. However, quantitative deviations will occur away from the transition point.
In the next version of the software, a mode for the correct handling of the Green's functions and susceptibilities to examine the accuracy of the approximation method will be implemented.
Finally, for convenience, a schematic flow of 
the RPA calculation is presented in Fig.~\ref{fig:rpa-flow}.

\section{Usage}
\label{Sec:Usage}

\subsection{Installing \hwave}\label{subsec:install}

\hwave is implemented in Python3 and requires several external libraries, such as
\textsc{NumPy}~\cite{harris2020array,numpy}, 
\textsc{SciPy}~\cite{2020SciPy-NMeth,scipy}, 
\textsc{tomli}~\cite{python-tomli}, 
and \textsc{requests}~\cite{python-requests}.
Since \hwave is registered to the Python Package Index (PyPI) repository~\cite{python-pypi},
users can install \hwave 
using the command-line tool \texttt{pip} as
\begin{verbatim}
  $ python3 -m pip install h-wave
\end{verbatim}
Additionally, the required libraries are installed.

The source code for \hwave is available in the GitHub repository~\cite{hwave-github}.
The users can download the zipped file from the release page, or clone the repository to
obtain the latest version.
After unpacking the zipped source files and changing the current directory to the top of the source tree, users can install \hwave by executing the following command.
\begin{verbatim}
  $ python3 -m pip install .
\end{verbatim}
Subsequently, \hwave and the required libraries are installed, and the executable file \texttt{hwave} is placed in the specified installation path.
Further details are provided in the installation section of the manual~\cite{hwave-manual}.

\subsection{Using \hwave}\label{subsec:usage}
\begin{figure}[t]
  \centering
  \includegraphics[width=.45\textwidth]{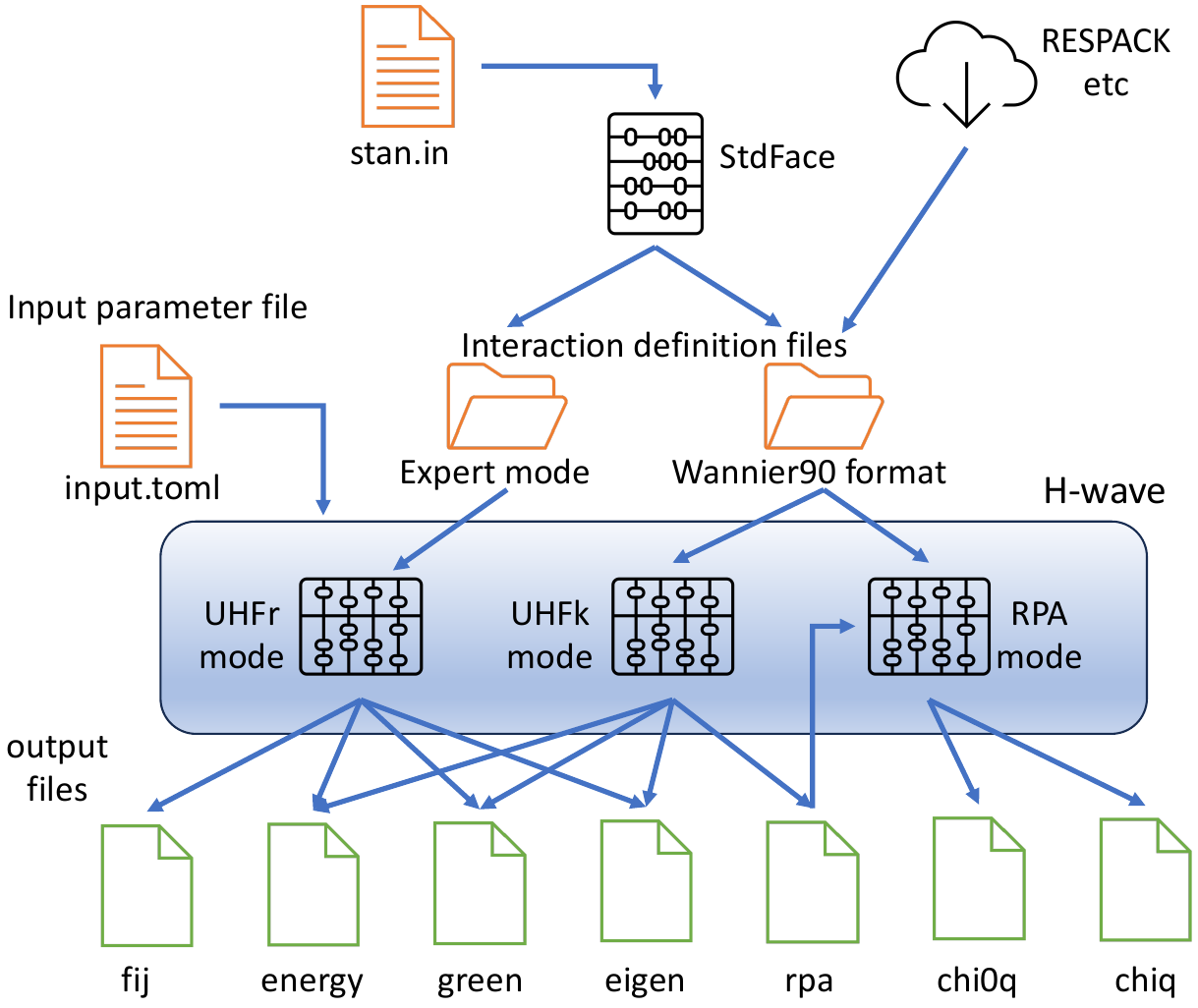}
  \caption{
    Schematic flow of calculations. First, the users prepare an input parameter file (\texttt{input.toml}) and a set of interaction definition files. The interaction definition files can be generated by \textsc{StdFace} from a simple definition file (\texttt{stan.in}), or the output of other software such as \textsc{RESPACK}\cite{Nakamura2021} can be used. Then, \hwave is run in a chosen mode. The results are provided as output in the files according to the input parameters.
  }
  \label{fig:calc_flow}
\end{figure}

To use \hwave, a \textsc{toml}-format file (e.g., \texttt{input.toml}) and input files that specify the parameters in the Hamiltonians have to be prepared.
Figure~\ref{fig:calc_flow} shows the schematic flow of the calculations.
In the \textsc{toml} file, the calculation parameters and the names of the input files for the Hamiltonian are specified.
For standard models such as the Hubbard model on a square lattice,
users can use the \textsc{StdFace} library~\cite{stdface} to generate the input files that specify the Hamiltonians.
Users can run \hwave by executing the following command.
\begin{verbatim}
  $ hwave input.toml
\end{verbatim}
When the calculations are complete, 
the results are provided as outputs to the files, which are specified by \texttt{input.toml}.

\subsubsection{Input parameter file}
The input parameter file is written in the \textsc{toml}-format~\cite{toml-format} text file.
In the \textsc{toml} format, the value of a parameter is specified in the form \texttt{parameter} = \texttt{value}, where the value can be a string, number, Boolean value, array, or table (associative array).
A set of parameters can be classified into a \textsc{toml} table structure and labeled by a name within a square bracket, which we call a \textit{section} hereafter.
An example of \texttt{input.toml} is presented below:
\begingroup
\setlength{\listingindent}{0pt}
\vskip2ex\hrule
\begin{listing}
[log]
  print_level = 1
  print_step = 10
[mode]
  mode = "UHFk"
[mode.param]
  2Sz = 0
  Ncond = 16
  IterationMax = 1000
  EPS = 12
  Mix = 0.5
  RndSeed = 123456789
  ene_cutoff = 1.0e+2
  T = 0.0
  CellShape = [ 4, 4, 1 ]
  SubShape = [ 2, 2, 1 ]
[file]
[file.input]
  path_to_input = ""
[file.input.interaction]
  path_to_input = "./"
  Geometry = "geom.dat"
  Transfer = "transfer.dat"
  CoulombIntra = "coulombintra.dat"
[file.output]
  path_to_output = "output"
  energy = "energy.dat"
  eigen = "eigen"
  green = "green"
\end{listing}
\hrule\vskip2ex
\endgroup

Important sections used in \texttt{input.toml}, such as
\texttt{[mode]}, \texttt{[mode.param]}, \texttt{[log]}, \texttt{[file.input]}, and 
\texttt{[file.output]} are discussed in this paper. 
Further details have been provided in the manual of \hwave~\cite{hwave-manual}.

\texttt{[mode]}---
In the \texttt{mode} section, 
the calculation mode is specified as
\begin{listing}
[mode]
mode = "UHFk"
\end{listing}
As aforementioned, \hwave implements three methods,
UHFA in real space (\texttt{mode="UHFr"}),
UHFA in wave-number space (\texttt{mode="UHFk"}),
and RPA (\texttt{mode="RPA")}.
During the UHFA calculations,
decision to include the Fock term can be specified using \texttt{flag\_fock}.
In the wave-number space UHFA and RPA calculations, \texttt{enable\_spin\_orbital} can be used to allow the hopping integrals that break the total $S_z$ conservation.
If the total $S_{z}$ is conserved, the calculation cost of RPA can be significantly reduced.
See Sec.~4.2.2 of the \hwave manual for the index rule of the orbitals and spins when the \texttt{enable\_spin\_orbital} is set to true.

\begin{figure}
\centering
\includegraphics[width=.4\textwidth]{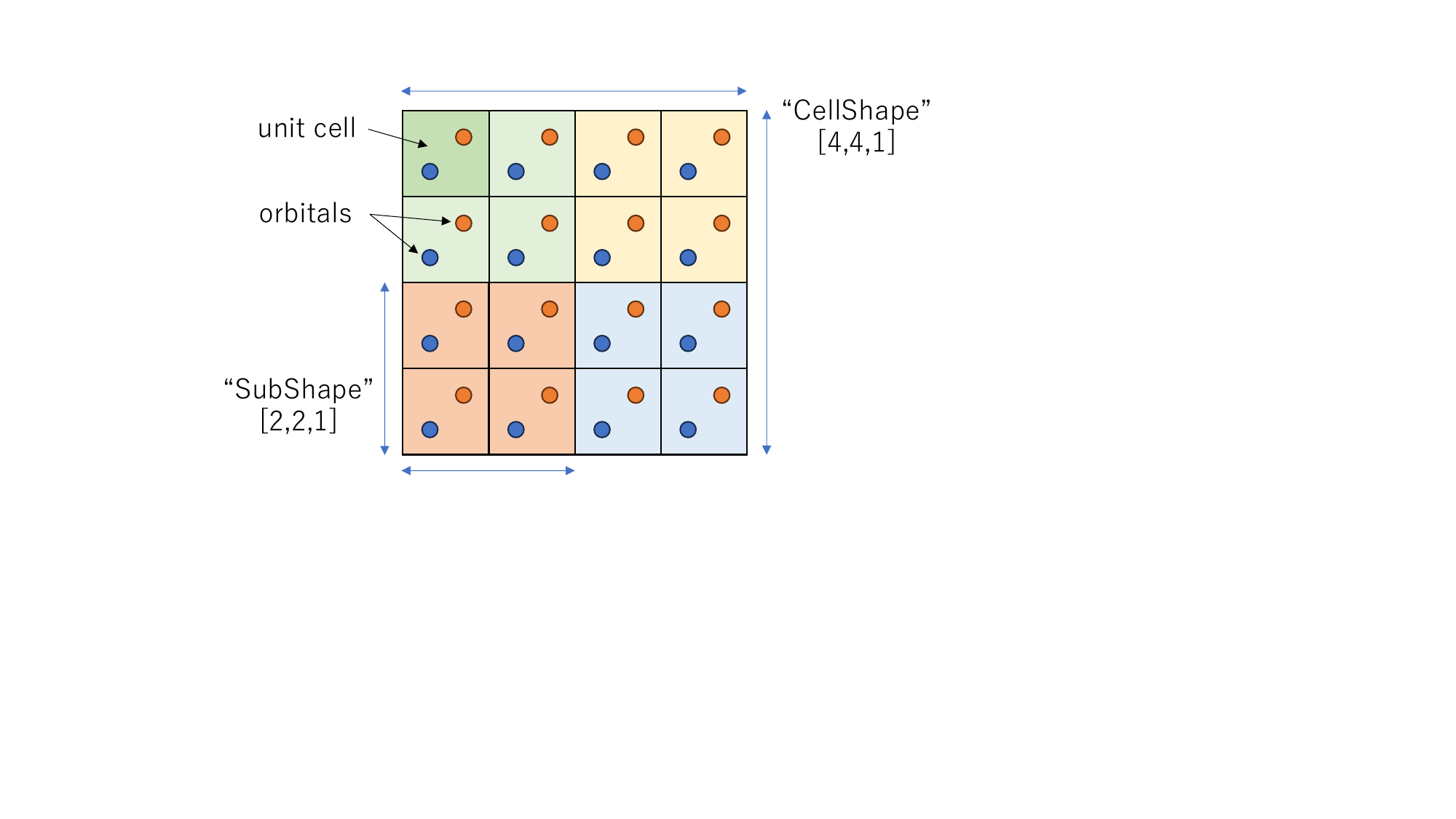}
\caption{The definitions of the overall lattice (CellShape) and the sublattice structure (SubShape) are depicted with examples.}
\label{fig:lattice}
\end{figure}

\texttt{[mode.param]}---In the \texttt{mode.param} section, the parameters concerning the calculation conditions can be specified. The lattice size (the number of sites) is specified using \texttt{CellShape} (\texttt{Nsite}). 
For the wave-number space UHFA and RPA calculations, a sub-lattice structure can be introduced using \texttt{SubShape}, as depicted in Fig.~\ref{fig:lattice}.
The temperature is denoted by \texttt{T}. 
The number of electrons is specified using \texttt{Ncond} or 
by \texttt{filling} ratio to the total number of states. 
In the UHFA calculations, the total spin can be specified using \texttt{2Sz}, or 
it can be left free. 

The mean-field approximation is iteratively solved using the convergence criterion given by the \texttt{EPS} parameter. For the RPA calculations, the number of Matsubara frequencies is specified using \texttt{Nmat}. 
For sufficiently large Matsubara frequency, the Green's function behaves like $1/i\omega_n$.
The high-energy tail of the Green's functions along the Matsubara frequencies can be improved by subtracting $a/i\omega_n$, which can be treated analytically. The coefficient of the subtraction term $a$ can be specified through \texttt{coeff\_tail}.
Basically, the default value $a=1$ works well, but for more precise work,  it may be better to select $a$ as $i\omega_{n_{\rm max}} G(i\omega_{n_{\rm max}})$.

\texttt{[log]}--- In the \texttt{log} section, the conditions of calculation logs can be specified. \texttt{print\_level} is used to specify the verbosity of the log outputs. 
In the UHFA calculations, the interval between the residue outputs during the iterations can be set using \texttt{print\_step}. 
When specified, the residues are written to the files given by \texttt{print\_check}.

\texttt{[file.input]}--- In the \texttt{file.input} section, the input files can be specified.
The files that define the Hamiltonians are specified in a separate sub-section \texttt{[file.input.interaction]}, as described in Sec.~\ref{sec:interactions}. 
In this section, the file name of the initial Green's function can be specified using \texttt{initial} parameter. 
In the RPA calculations, the interaction term approximated using the UHFA can be considered for the initial configuration.
The coefficient of the modified hopping term $\widetilde{\mathcal{H}_0} = \sum \widetilde{t_{ij}^{\alpha\beta}} c_{i\alpha}^\dagger c_{j\beta}^{\mathstrut} + \textit{H.c.}$, where
\begin{equation}
  \widetilde{t_{ij}^{\alpha\beta}} = t_{ij}^{\alpha\beta}
  + \sum_{k,\gamma,\gamma^\prime} W_{ik}^{\gamma\gamma^\prime,\alpha\beta} \langle c_{k\gamma^\prime}^\dagger c_{k\gamma}^{\mathstrut} \rangle\,\delta_{ij},
\end{equation}
can be read from a file specified by \texttt{trans\_mod}.
The input file can be generated through a wave-number space UHFA calculation using the \texttt{rpa} output option.
An option for using the irreducible susceptibility $\chi^{(0)}(\vec{q})$ 
specified by \texttt{chi0q}, which is obtained from the former calculation, is available for the calculation of the susceptibility matrix.

\texttt{[file.output]}--- In the \texttt{file.output} section,
the output files for the calculation results can be specified.
The items as output for each calculation mode are described in Sec.~\ref{sec:usage:output}.

\subsubsection{Interaction definition files}
\label{sec:interactions}
\hwave supports a set of interaction definition files that define the Hamiltonian. They are specified in the \texttt{[file.input.interaction]} section of the parameter file with keywords and associated file names. These files consist of geometry information (labeled as \texttt{Geometry}), transfer integral (labeled as \texttt{Trans} for UHFr or \texttt{Transfer} for UHFk and RPA), and two-body interaction terms with names based on the convention adopted in other software for quantum lattice models, \HPhi~\cite{KAWAMURA2017180,ido2023update} and \mVMC~\cite{Misawa2019}, as listed in Table~\ref{table:interactions}.
\begin{table}[htb]
  \caption{%
Keywords for the interaction types and their descriptions are summarized.
$c_{i\alpha\uparrow(\downarrow)}^\dagger$ and $c_{i\alpha\uparrow(\downarrow)}$ denote the creation and annihilation operators, respectively, of electrons at site $i$ and orbital $\alpha$ with spin-up (spin-down).
$N_{i\alpha\uparrow(\downarrow)} = c_{i\alpha\uparrow(\downarrow)}^\dagger c_{i\alpha\uparrow(\downarrow)}$,
and $N_{i\alpha} = N_{i\alpha\uparrow} + N_{i\alpha\downarrow}$.
$S^\nu_{i\alpha} = \frac{1}{2} \sum_{\sigma, \sigma'} c^\dagger_{i\alpha \sigma} \sigma^\nu_{\sigma \sigma'} c_{i\alpha \sigma'}$ where $\vec{\sigma}^k$ is the Pauli matrix, and $\vec{\sigma}^{\pm} = \frac{1}{2}(\vec{\sigma}^x \pm i\vec{\sigma}^y)$.
}
  \begingroup
  \renewcommand{\baselinestretch}{1.2}
  \small
  \begin{tabular}{p{.12\textwidth}p{.33\textwidth}}
    \toprule
    Keyword & Description \\
    \midrule
    \texttt{Transfer} (or \texttt{Trans}) &
    Transfer term denoted by $T_{i\alpha j\beta}\, c_{i\alpha\sigma}^\dagger c_{j\beta\sigma}^{\phantom{\dagger}}$ (\texttt{spin\_orbital} = \texttt{False}),  or $T_{i(\alpha , \sigma) j (\beta, \sigma')}\, c_{i(\alpha,\sigma)}^\dagger c_{j(\beta ,\sigma')}^{\phantom{\dagger}}$ (\texttt{spin\_orbital} = \texttt{True}) \\
    \texttt{InterAll} &
    A general two-body interaction term of the form $I_{ijkl\sigma_1\sigma_2\sigma_3\sigma_4}\, c_{i\sigma_1}^\dagger c_{j\sigma_2}^{\phantom{\dagger}} c_{k\sigma_3}^\dagger c_{l\sigma_4}^{\phantom{\dagger}}$ (only for UHFr), \\
    \texttt{CoulombIntra} &
    On-site Coulomb interaction denoted by $U_{i\alpha}\,N_{i\alpha\uparrow} N_{i\alpha\downarrow}$ \\
    \texttt{CoulombInter} &
    Off-site Coulomb interaction denoted by $V_{i\alpha j\beta}\,N_{i\alpha} N_{j\beta}$ \\
    \texttt{Hund} &
    Hund coupling interaction denoted by $J_{i\alpha j\beta}^\text{Hund}\, ( N_{i\alpha\uparrow} N_{j\beta\uparrow} + N_{i\alpha\downarrow} N_{j\beta\downarrow} )$ \\
    \texttt{Ising} &
    Ising interaction denoted by $J_{i\alpha j\beta}^\text{Ising}\, S_{i\alpha}^{z} S_{j\beta}^{z}$ \\
    \texttt{Exchange} &
    Exchange interaction denoted by $J_{i\alpha j\beta}^\text{Ex}\, S_{i\alpha}^{+} S_{j\beta}^{-}$ \\
    \texttt{PairLift} &
    The interaction denoted by $J_{i\alpha j\beta}^\text{PairLift}\, c_{i\alpha\uparrow}^\dagger c_{i\alpha\downarrow}^{\phantom{\dagger}} c_{j\beta\uparrow}^\dagger c_{j\beta\downarrow}^{\phantom{\dagger}}$ \\
    \texttt{PairHop} &
    The interaction denoted by $J_{i\alpha j\beta}^\text{PairHop}\, c_{i\alpha\uparrow}^\dagger c_{j\beta\uparrow}^{\phantom{\dagger}} c_{i\alpha\downarrow}^\dagger c_{j\beta\downarrow}^{\phantom{\dagger}}$ (only for \texttt{UHFr} mode) \\
    \bottomrule
  \end{tabular}
  \endgroup
  \label{table:interactions}
\end{table}
For the RPA calculations, an external field to the one-body interaction term can be introduced using the keyword \texttt{Extern}. 
The form of the term added to the transfer term $T_{\alpha\sigma,\beta\sigma^\prime}(r)$ reads
\begin{equation}
  T_{\alpha\sigma,\beta\sigma^\prime}(r) \to T_{\alpha\sigma,\beta\sigma^\prime}(r) + h (\bm{\sigma}^z)_{\sigma\sigma^\prime} B_{\alpha\beta}(r),
\end{equation}
where $B_{\alpha\beta}(r)$ is the external field given in the file specified by \texttt{Extern} keyword, $h$ is the coefficient specified by \texttt{extern\_coeff} parameter, and $\bm{\sigma}^z$ is the $z$-component of the Pauli matrix.

The file format of the interaction definition files for the UHFr calculations follows the 
\textit{Expert-mode} format of \HPhi, which is a text-based format
with header lines, followed by content in which 
the indices $i$ and $j$ and the value of the coefficient, for example, $J_{ij}$, 
are listed for the interaction term.
Further details have been provided in the manuals of \HPhi~\cite{hphi-manual} and \mVMC~\cite{mvmc-manual}.

For the UHFk and RPA calculations, a \textsc{Wannier90}-like format~\cite{Pizzi2020}
is adopted for the interaction definition because translational symmetry is assumed in these methods.
In this format, the coefficient of the interaction term $J_{i\alpha j\beta} = J_{\alpha\beta}(r_{ij})$ can be expressed by the components of the three-dimensional translation vector $r_{ij}$, 
the orbital indices $\alpha$ and $\beta$, and the value in each line. 
\hwave allows zero components to be omitted.
A few examples of the interaction definition files for the UHFk calculations are presented in the following.

\noindent
The \texttt{Transfer} term can be expressed as
\begingroup
\small
\setlength{\listingindent}{0pt}
\vskip2ex\hrule
\begin{listing}
Transfer in Wannier90-like format for uhfk
1
9
 1 1 1 1 1 1 1 1 1
  -1    0    0    1    1  -1.000   0.000
   0   -1    0    1    1  -1.000   0.000
   0    1    0    1    1  -1.000   0.000
   1    0    0    1    1  -1.000   0.000
\end{listing}
\hrule\vskip2ex
\endgroup
\noindent
The on-site Coulomb interaction (\texttt{CoulombIntra}) can be expressed as
\begingroup
\small
\setlength{\listingindent}{0pt}
\vskip2ex\hrule
\begin{listing}
CoulombIntra in Wannier90-like format for uhfk
1
1
 1
   0    0    0    1    1   4.000   0.000
\end{listing}
\hrule\vskip2ex
\endgroup
\noindent
The geometry definition (\texttt{Geometry}) can be expressed as
\begingroup
\small
\setlength{\listingindent}{0pt}
\vskip2ex\hrule
\begin{listing}
  1.000   0.000   0.000
  0.000   1.000   0.000
  0.000   0.000   1.000
1
    0.000e+00     0.000e+00     0.000e+00
\end{listing}
\hrule\vskip2ex
\endgroup
\noindent
The \textit{file specifications} section of the \hwave manual~\cite{hwave-manual} can be referred to for details on the file formats.

Such files can be generated from a simple input using the \textsc{StdFace} library~\cite{stdface}.
An example of the input file (\texttt{stan.in}) has been provided
\begingroup
\small
\setlength{\listingindent}{0pt}
\vskip2ex\hrule
\begin{listing}
model     = "Hubbard"
lattice   = "square"
W         = 4
L         = 4
t         = 1.0
U         = 4.0
calcmode  = "uhfk"
exportall = 0
\end{listing}
\hrule\vskip2ex
\endgroup

\subsubsection{Output}\label{sec:usage:output}
\hwave outputs the calculation results to the files according to the settings specified in the \texttt{[file.output]} section of the input parameter file. A brief overview of the main parameters in \texttt{[file.output]} is discussed.

In real space UHFA calculations, the following keywords can be used.
\begin{itemize}
\item \texttt{energy}: The total energy and its components for each interaction term, the number of electrons, and the value of $S^z$ are written to a text file in the \texttt{item=value} format.
\item \texttt{eigen}: The eigenvalues and eigenvectors are provided as outputs in \textsc{NumPy} zip files~\cite{numpy-zip-format}.
For a fixed $S^z$, the spin-up and spin-down components are stored in separate files with \texttt{spin-up} or \texttt{spin-down} prefix. Otherwise, the eigenvalues and eigenvectors are provided as outputs in a single file with a prefix \texttt{sz-free}.
\item \texttt{green}: The Green's function is provided as an output in a text file with indices specified in the \texttt{OnebodyG} file.
\item \texttt{fij}: In real-space UHFA calculations, the coefficients of the pair-product wave functions $f_{ij}$, which are equivalent to the converged one-body wave functions, are provided as outputs. They are written to a text-format file specified by \texttt{fij}. Further details on the relationship between the HFA solutions and the pair-product wave functions have been provided in Ref.~\cite{Misawa2019}.
\end{itemize}

In the \texttt{UHFk} calculations, the following keywords can be used.
\begin{itemize}
\item \texttt{energy}: The value of the total energy and its components for each interaction term, 
the number of electrons, and the value of $S^z$ are written to a text file in the \texttt{item=value} format.
An example output of \texttt{energy} is shown below.
\begingroup
\small
\setlength{\listingindent}{0pt}
\vskip2ex\hrule
\begin{listing}
Energy_Total        = -12.56655452028271
Energy_Band         = -4.507491115073952
Energy_CoulombIntra = -8.059063405208757
NCond   = 15.999999999999996
Sz      = -7.239267874048494e-08
\end{listing}
\hrule\vskip2ex
\endgroup
\item \texttt{eigen}: The eigenvalues and eigenvectors are provided as outputs in a \textsc{NumPy} zip file.
\item \texttt{green}: The Green's function is provided as an output in a \textsc{NumPy} zip file.
\item \texttt{rpa}: The modified Transfer term representing the approximated one-body Hamiltonian is provided as an output in a \textsc{NumPy} zip file for an initial value of the RPA calculation.
\end{itemize}

In the RPA calculations, the following keywords can be used.
\begin{itemize}
\item \texttt{chi0q}: The irreducible susceptibility matrix $\chi^{(0)}(\vec{q}, i\nu_n)$ is provided as an output in a \textsc{NumPy} zip file.
\item \texttt{chiq}: The susceptibility matrix $\chi(\vec{q}, i\nu_n)$ is provided as an output in a \textsc{NumPy} zip file.
\end{itemize}

\section{Applications}
\label{Sec:Examples}
In this section, three examples of \hwave have been introduced, which include methods to 
calculate a ground-state phase diagram at zero temperature~(zero-temperature UHFk), finite-temperature 
physical quantities (finite-temperature UHFk),
and charge and spin susceptibilities at finite temperatures (RPA).
Tutorials on \hwave have been uploaded to a repository~\cite{gitlab}, 
which includes several other samples of \hwave and examples presented in this paper.

\subsection{Ground state of the extended Hubbard model on a square lattice}
\begin{figure}[t]
    \centering
    \includegraphics[width=70mm]{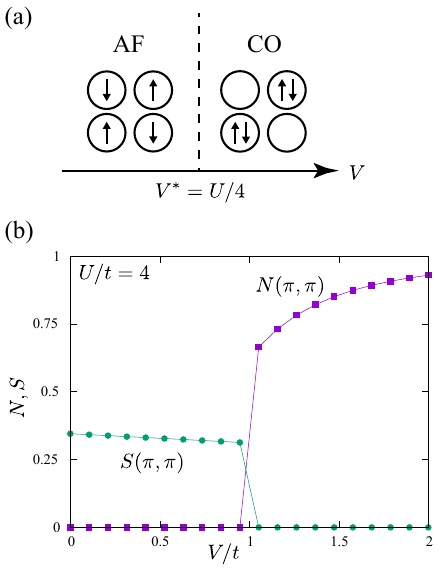}
    \caption{
    (a) Expected phase diagram of the extended Hubbard model on the square lattice at half-filling in the strong coupling limit.
    (b)
    $V$ dependencies of the staggered magnetization $S(\pi,\pi)$ (green circles) and the staggered charge density $N(\pi,\pi)$ (blue squares) of the extended Hubbard model on the square lattice with $U=4$ 
    at half-filling calculated by the UHFk mode of \hwave.
    A first-order phase transition between the AF phase and the CO phase occurs around $V/U=1/4$.
    }
    \label{fig:cdwsdw}
\end{figure}

As a simple application of the UHFk calculation at zero temperature,
the ground states of the extended Hubbard model on a square lattice at half-filling were considered.
In the extended Hubbard model, a first-order quantum phase transition is expected from 
the antiferromagnetic (AF) phase to 
the charge-ordered (CO) phase by changing the off-site Coulomb interactions.
The extended Hubbard model is defined as
\begin{multline}
\mathcal{H} = 
-t\sum_{\ev*{i,j}}\sum_{\sigma = \uparrow\downarrow} \left[ c_{i\sigma}^\dagger c_{j\sigma}^{\mathstrut} + c_{j\sigma}^\dagger c_{i\sigma}^{\mathstrut} \right] \\
+ U\sum_i N_{i\uparrow}N_{i\downarrow} + V\sum_{\ev{i,j}}N_iN_j, \label{eq:extendedHubbard}
\end{multline}
where $\ev*{i,j}$ means the nearest neighbor pairs on the square lattice, $N_{i\sigma} = c^\dagger_{i\sigma}c_{i\sigma}^{\mathstrut}$ represents the number operator of the spin $\sigma$ at the site $i$, $N_i = N_{i\uparrow}+N_{i\downarrow}$, and $U$ and $V$ denote the onsite and offsite Coulomb repulsion.
The hopping constant $t$ was used as the unit of energy. (Hereafter, $t=1$.) 
The case of half-filling was considered, in which the number of electrons was taken to be equal to the number of sites.
Two phases are expected to appear as the ground states (Fig.~\ref{fig:cdwsdw}(a)).
For $V/U \lesssim 1/4$, the AF phase where the neighboring spins align in opposite directions to each other emerges.
The AF phase is characterized by the staggered magnetization defined as
$S(\mathbf{Q}) = \sum_i S_i^z \exp(-i \mathbf{Q}\cdot\mathbf{r}_i)/N_{\text{cell}}$
with $\mathbf{Q}=(\pi,\pi)$, where
$S_i^z = (N_{i\uparrow} - N_{i\downarrow})/2$ and the summation runs over sites in the unit cell, 
and $N_{\text{cell}}$ is the number of sites in the unit cell.
For $V/U \gtrsim 1/4$, the CO phase becomes the ground state where the doubly occupied sites are arranged in a checkerboard pattern.
The CO phase is characterized by the staggered density defined as $N(\mathbf{Q}) = \sum_i (N_{i\uparrow}+N_{i\downarrow})\exp(-i \mathbf{Q}\cdot \mathbf{r}_i)/N_{\text{cell}}$.

Figure~\ref{fig:cdwsdw} (b) shows the $V$ dependence of $S(\pi,\pi)$ and $N(\pi,\pi)$.
As expected, the first-order phase transition
between the AF and the CO phases was observed at $V/t = 1 ( = U/4t)$.
A sample script is available in \texttt{samples/UHFk/CDW\_SDW} directory of the tutorial repository.
By executing the script \texttt{run.py}, the results shown in Figure~\ref{fig:cdwsdw}~(b) can be reproduced.

\subsection{Finite-temperature properties of the Hubbard model on a cubic lattice}

\hwave supports finite-temperature mean-field calculations. 
In this section, a typical simple example involving the estimation of ${\rm N\acute{e}el}$ temperature for 
the Hubbard model on a cubic lattice at half-filling has been discussed. The model was defined as
\begin{align}
  \mathcal{H} &= \sum_{\bm{r}} \left[ -t \sum_{\nu, \sigma} \sum_{\bm{\delta}=\pm \bm{a}_{\nu}} c_{\bm{r}\sigma}^\dagger c_{\bm{r}+\bm{\delta}, \sigma} + U N_{\bm{r}\uparrow} N_{\bm{r}\downarrow} \right],
\end{align}
where $\bm{a}_\nu$ denotes the translation vector along the $\alpha$-axis, and $\nu = x$, $y$, and $z$.

\begin{figure}[t]
    \centering
    \includegraphics[width=70mm]{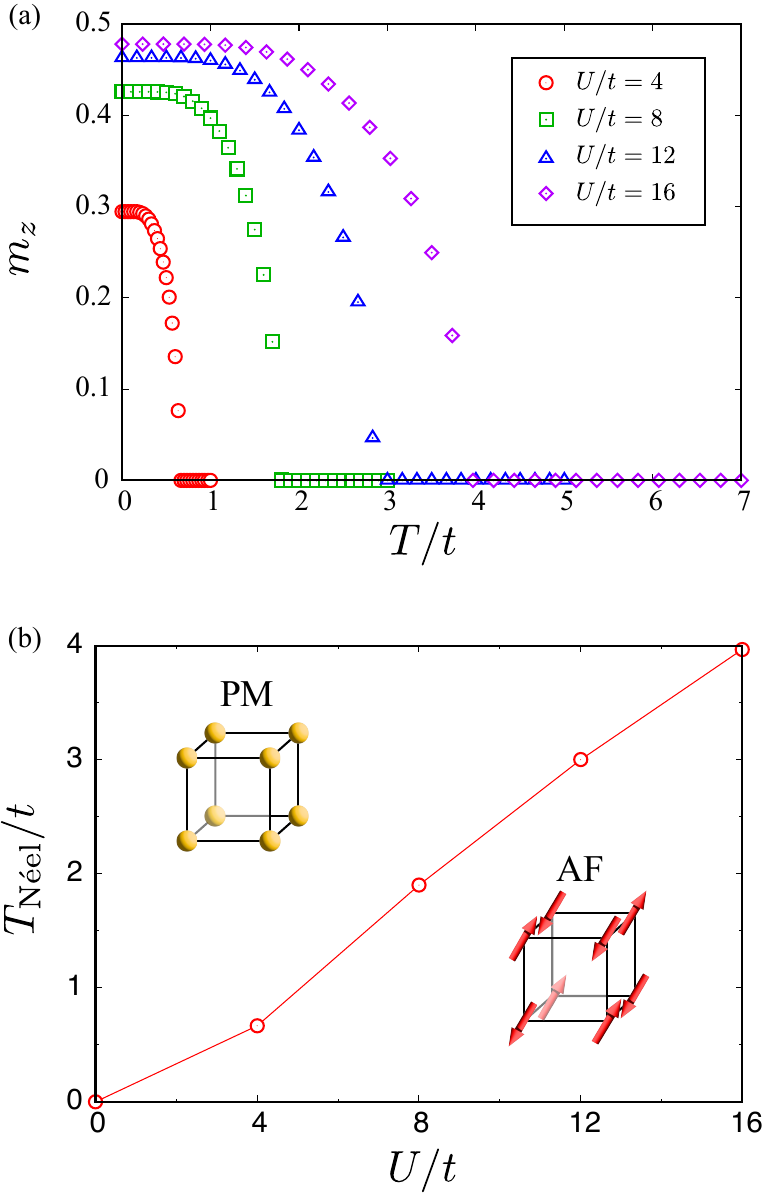}
    \caption{Magnetic properties of the Hubbard model on the cubic lattice at half-filling for $L=12$. (a) Temperature dependence of the magnetic moment along the $z$-axis. Red circles, green squares, blue triangles, and purple diamonds denote the mean-field results for $U/t=4$, $8$, $12$, and $16$, respectively. (b) Interaction dependence of the ${\rm N\acute{e}el}$ temperature $T_{\rm N\acute{e}el}$. Red circles denote the mean-field results. Insets represent schematics of paramagnetic (PM) and anti-ferromagnetic (AF) states.}
    \label{fig:Hubbard_FiniteT_L12}
\end{figure}

Figure~\ref{fig:Hubbard_FiniteT_L12}(a) shows the dependency of magnetization on temperature for the Hubbard model on 
a cubic lattice at half-filling. 
The magnetization presented along the $z$ axis was defined as
\begin{align}
  m_z &= \frac{1}{2N_\text{site}} \sum_{\bm{r}} \left| \ev*{n_{\bm{r}\uparrow}- n_{\bm{r}\downarrow}} \right|.
\end{align}
The number of sites $N_\text{site}$ was denoted as $N_\text{site} = L^3$, where $L$ denotes 
the linear dimensions of a cubic structure.
The magnetization was observed to be finite at 
the critical temperature $T_{\rm N\acute{e}el}$. 
For this study, $T_{\rm N\acute{e}el}$ was regarded as the lowest temperature at which $m_z$ had a value lesser than $10^{-4}$.

Figure \ref{fig:Hubbard_FiniteT_L12}~(b) shows the interaction dependence of ${\rm N\acute{e}el}$ temperature $T_{\rm N\acute{e}el}$. The results suggested that $T_{\rm N\acute{e}el}$ monotonically increased with the interaction $U/t$. However, this tendency was a well-known artifact of the HFA.
For a strong-coupling limit, $T_{\rm N\acute{e}el}$ is expected to decrease with increase in $U/t$ \cite{Kakehashi1987, Rohringer2011, Mukherjee2014} 
because it is governed by super-exchange interactions $J=4t^2/U$. 
To correctly reproduce such a tendency, 
many-body correlations beyond the HFA must necessarily be included.

\subsection{Charge and spin susceptibilities of the extended Hubbard model on a square lattice}
In this sub-section, charge and spin susceptibilities of the extended Hubbard model on a square lattice defined in Eq.~\eqref{eq:extendedHubbard} using the RPA are discussed. In order to reproduce the numerical results shown in Fig.~1 of Ref.~\cite{Kobayashi2004b}, the electron filling was set as $3/4$ and $T=0.01$. The cell size was set to $L_x = L_y = 128$.  
Sample files and scripts were used from the \verb|samples/RPA/kobayashi_2004| directory of the tutorial repository.

\begin{figure}
    \centering
    \includegraphics[width=80mm]{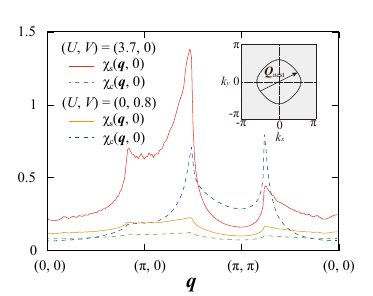}
    \caption{Spin and charge susceptibilities of the extended Hubbard model at $T=0.01$. The interaction parameters are set as $(U, V) = (3.7, 0)$ and $(0, 0.8)$, respectively. The inset shows the Fermi surface (solid line) and the nesting vector ${\bm Q}_{\rm nest}\equiv(\pi, \pi/2)$.}
    \label{fig:rpa-chi}
\end{figure}

Figure~\ref{fig:rpa-chi} shows spin and charge susceptibilities $\chi_s ({\bm q}, 0)$, $\chi_c ({\bm q}, 0)$ at $(U, V) = (3.7, 0)$ and $(0, 0.8)$, respectively, where $\chi_s ({\bm q}, 0)$ and $ \chi_c ({\bm q}, 0) $ are defined as $1/2 \sum_{\sigma\sigma'}\sigma\sigma'\chi_{\sigma \sigma'}({\bm q}, 0)$ with $\sigma = +(-)$ for $\uparrow (\downarrow)$ and $ 1/2\sum_{\sigma\sigma'}\chi_{\sigma \sigma'}({\bm q}, 0)$, respectively. 
At $(U, V) = (3.7, 0)$, $\chi_s({\bm q}, 0)$ showed a sharp peak around ${\bm Q}_{\rm nest}=(\pi, \pi/2)$ due to the nesting condition as shown in the inset of Fig.~\ref{fig:rpa-chi}. This peak developed with increasing $U$ and diverged when the spin-density-wave transition occurred. 
At $(U, V) = (0, 0.8)$, $\chi_c({\bm q}, 0)$ exhibited sharp peaks around $(\pi, \pi/2)$ and $(\pi/2, \pi/2)$. The first peak originated from the nesting condition, whereas the second peak originated from the anisotropy of $V_{\bm q}$, which had a negative peak at $(\pi, \pi)$, in addition to the nesting conditions. The latter peak developed with an increase in $V$ and diverged when a charge-density wave (or charge ordering) transition occurred. 
The obtained results aligned with those shown in Fig.~1 of Ref.~\cite{Kobayashi2004b}.

\section{Summary}\label{Sec:Summary}

In this paper,  we introduced \hwave, which can perform the unrestricted Hartree--Fock approximation (UHFA) and the random phase approximation (RPA). 
UHFA and RPA in wave-number space enable numerical analyses of electron correlation effects in periodic electron systems within fluctuations up to the first order at a low computational cost. 
\hwave is open-source software with simple and flexible user interface.
Users can execute UHFA and RPA for widely studied quantum lattice models such as the Hubbard model by preparing only one input file with less than ten lines. 
Furthermore, the input files of one- and two-body interactions describing effective models can be imported from first principles calculations using the \textsc{Wannier90} format such as \textsc{RESPACK}. This enables a high-throughput connection between the first principles calculations and \hwave. The resulting files can be used as input files for softwares that handle strong correlation effects, such as \mVMC~\cite{Misawa2019}.
\hwave is included in MateriApps LIVE!~\cite{MALIVElink,MALIVE}, an environment for computational materials science, and MateriApps Installer~\cite{MALIVE,MAINSTALLERlink}, a collection of scripts for installing materials science calculation software. These provide the environment to easy application of \hwave.

In the future, we plan to develop the following functions based on the experience gained from our previous works; (a) function to calculate the quantities corresponding to the experimentally measured dynamic susceptibilities, such as magnetic susceptibility and spin-lattice relaxation rate for nuclear magnetic resonance and conductivity~\cite{doi:10.1143/JPSJ.12.570, doi:10.1143/JPSJ.18.516, doi:10.1143/JPSJ.59.2508, Ohki2022}, (b) function to solve the linear Eliashberg equation evaluating the superconducting transition temperature and order parameters by considering pairing interactions mediated by charge and spin fluctuations~\cite{Eliashberg, Kobayashi2004a,Kobayashi2004b,yoshimi2007}, and (c) function to treat
approximations beyond RPA such as fluctuation exchange approximation (FLEX) and vertex corrections~\cite{BICKERS1989206, PhysRevLett.62.961, yoshimi2009,yoshimi2011, yoshimi2012}.

\section*{Acknowledgments}
\hwave is developed under the support of ``Project for Advancement of Software Usability in Materials Science'' in the fiscal year 2022 by The Institute for Solid State Physics, The University of Tokyo.
This work was supported by JSPS KAKENHI Grant Numbers 21H01041 and 22K03526.
\bibliographystyle{elsarticle-num}
\bibliography{main.bib}

\end{document}